%% file: mon_spec.tex
\documentstyle[art11]{article}

\newlength{\jumpback}
\def\lbr	{\mathord{l_{\hbox{\small br}}}}

\title{Monopole spectra in non--Abelian gauge theories.}

\author{\underline{A. Hart} and M. Teper\thanks{e--mails: harta@thphys.ox.ac.uk,
teper@thphys.ox.ac.uk}
\\
{\small\sl Theoretical Physics, Oxford University, 1, Keble Road, Oxford, OX1
3NP, 
United Kingdom.}}
       
\date{June 25, 1996}
\begin	{document}
\maketitle
\begin	{abstract}
\noindent
We study the continuum limit of the length spectrum of magnetic
monopole structures found after various Abelian projections of pure
gauge $SU(2)$, including the maximally Abelian gauge. 
We comment on Gribov copies, and measurements of the string tension.
\end{abstract}

\vfill
Oxford Preprint Number: {\em OUTP--96--36--P}
\hfill
hep-lat/9606022

\newpage
\section{Introduction.}

The idea of 't Hooft 
\cite{tHo}
that the long range degrees of freedom in non--Abelian gauge theories
might be monopoles in suitable Abelian projections (AP's) of the
fields may be realised in the maximally Abelian gauge (MAG)
\cite{MAG},
where the non--Abelian string tension appears to be reproduced by the
monopole currents (`monopole dominance', e.g. 
\cite{stack}).

Such behaviour seems limited to a small class of AP's; why this
includes the MAG is unresolved.  The MAG is plagued by Gribov copies
which differ in long--range quantities such as the string tension
\cite{hart_teper,Balinew}
and whose proper treatment is not known.

We seek to address the above issues by examining the continuum limit of
the conserved magnetic currents identified
\cite{DeGT}
after various AP's; although AP is strictly only concerned with
reproducing the long
distance physics, monopole dominance can only be taken seriously if
the monopoles behave as physical objects on most length scales.
We use this as a criterion for favouring the MAG
(Sect.~\ref{sect_mon_spec}) over other gauges
(Sect.~\ref{sect_other_grib}), where we also comment on the Gribov
ambiguity.

Lattice units are used, unless physical ones are specified which are
in terms of the (non--Abelian) string tension
\cite{michtep}.
The latter is an unambiguous, non--perturbative definition.

\section	{The Monopole Loop Spectrum.}

\label	{sect_mon_spec}

In four dimensions, magnetic 4--current 
is conserved, and occurs in
closed `clusters'. Typically only a few~\% of the dual links carrying
magnetic current carry current greater than unity, so small clusters
routinely take the form of a single, non--intersecting `loop' of
current. 

Larger clusters can be resolved into a set of intersecting loops,
but there is no unique solution to this. In
this work, one such solution was selected randomly. The effect of this
will be discussed in Sect.~\ref{ssect_tail}.

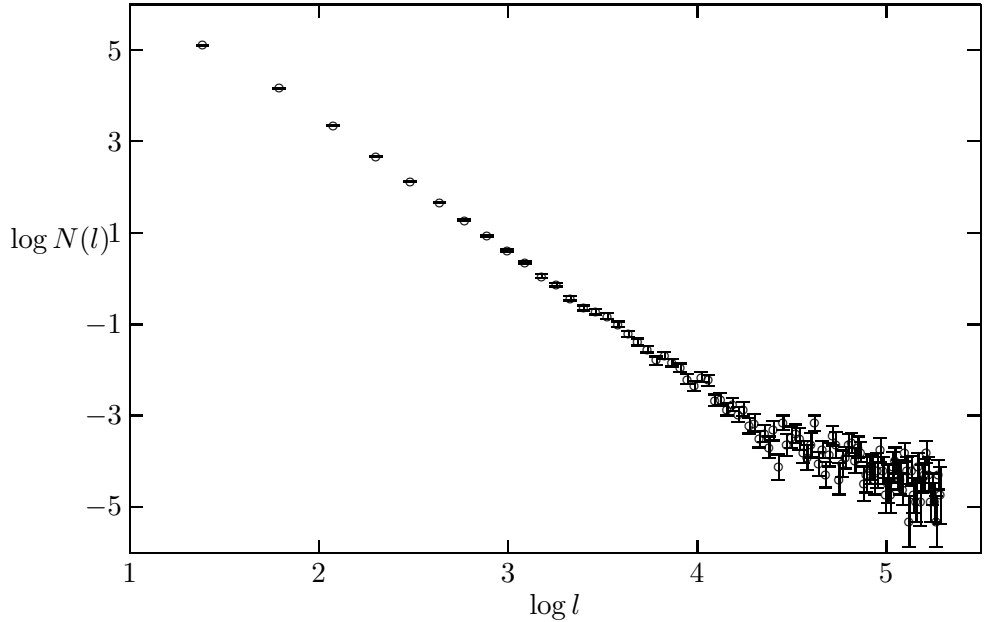
\begin	{figure}[t]
\begin	{flushright}
\leavevmode
\input	{plot_spec2}
\end	{flushright}

\vspace	{\jumpback}
\caption{The monopole loop spectrum at $\beta=2.4$ on a $14^4$ lattice.}
\label	{fig_plot_spec}

\vspace	{\jumpback}
\end 	{figure}

Fig.~\ref{fig_plot_spec} is a typical example of a `loop spectrum', a
plot of the ensemble--average number of monopole loops, $N(l)$, of
length, $l$ in the MAG.

The spectrum can be split into two parts; small
loops follow a clear power law dependence. There is an identifiable
point, $\lbr$, where this breaks down and there is an increase in
statistical errors. The number of large loops in the `tail' is
enhanced above the power law value.

\subsection	{The Power Law}

\label		{ssect_pow_law}

Loops larger than the plaquette
have lengths well described by a power law:
$$
N(l) = \frac{C}{l^\gamma}
$$
where $C$ is independent of $l$. 
Assuming that the power law holds for loops of arbitrary size (see
Sect.~\ref{ssect_tail}), $\gamma \geq 2$ ensures a finite total
current density.  
To generate a string tension there must be
sufficient large monopole loops to disorder the largest Wilson loops,
and $\gamma \leq 3$.
(A more model dependent argument \cite{diakonov}
requires $\gamma = 3$, based on the correspondence between instantons
and monopole loops, e.g.
\cite{hart_teper_pub}.)
 
Fig.~\ref{fig_plot_gamm} plots $\gamma$ against the lattice side
length in physical units for three values of $\beta_{SU(2)}$; a factor
two change in lattice spacing. $\gamma$ is within the above bounds and
consistent with a single value ${\overline \gamma} = 2.87(1)$, .

\begin	{figure}[t]
\begin	{flushright}
\leavevmode
\input{plot_cp2}
\end	{flushright}

\vspace{\jumpback}
\caption{Scaling of the spectrum density $c_p$. Key: $+$ -
$\beta=2.3$, $\times$ - $\beta=2.4$, $\star$ - $\beta=2.5$}
\label	{fig_plot_cp}

\vspace{\jumpback}
\end 	{figure}
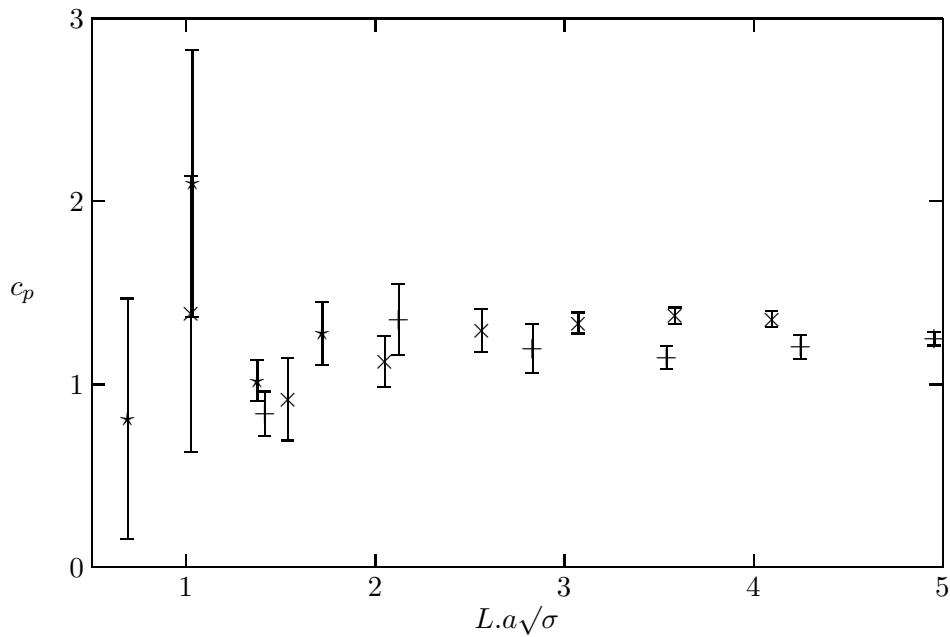

As we approach the continuum limit the number of monopole loops of a
given {\em physical} length,{\tt l}, per unit {\em physical} volume
$$
n(\hbox{\tt l}) = \frac{c_p}{\hbox{\tt l}^{\gamma}}
\hbox{\hspace{2em}}
c_p = \frac{C}{L^4(a\surd\sigma)^{4-\gamma}}
$$
also scales and $c_p$ remains near constant in 
Fig.~\ref{fig_plot_cp}
against changes in the lattice size, $L$, and the string tension,
$a\sigma$, with a mean value $\overline{c_p} = 1.27(2)$. Note
$\gamma$ implies a non--integer dimension for $c_p$.

\subsection	{Breakdown of the Power Law.}

\label		{ssect_break}

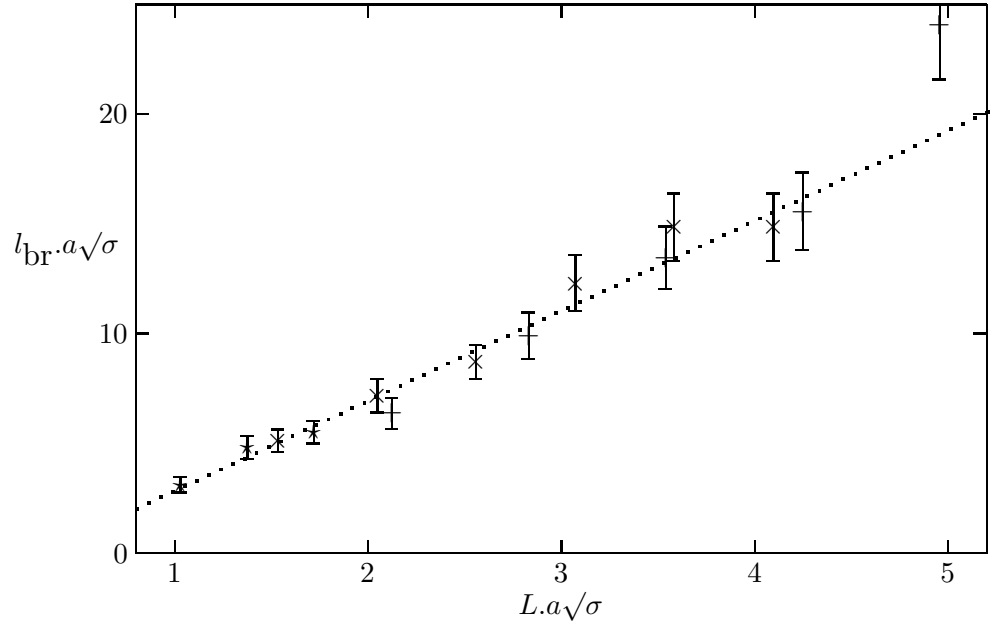
\begin	{figure}[t]
\begin	{flushright}
\leavevmode
\input	{plot_break_phys2}
\end	{flushright}

\vspace{\jumpback}
\caption{The breakdown point of the power law. Key: $+$ -
$\beta=2.3$, $\times$ - $\beta=2.4$, $\star$ - $\beta=2.5$}
\label	{fig_plot_break}

\vspace	{\jumpback}
\end 	{figure}

On a given lattice the power law
breaks down for the long loops at $l \simeq \lbr$,
which we find to be proportional to the lattice side length in
Fig.~\ref{fig_plot_break}.

The largest lattices are usually considered reasonable for extracting
continuum physics in terms of finite volume and spacing effects, and
the proportionality shows no sign of ceasing on larger lattices,
suggesting the linear relation continues and the entire monopole loop
spectrum follows the power law in the infinite volume limit.

\subsection	{The Tail of the Spectrum.}

\label		{ssect_tail}

The total current density scales less well 
\cite{magrev}
with the string tension than does the power law
describing the smaller loops containing around 50~\% of the
current. The non-scaling behaviour comes from the very largest loops.
There are two extreme possibilities:

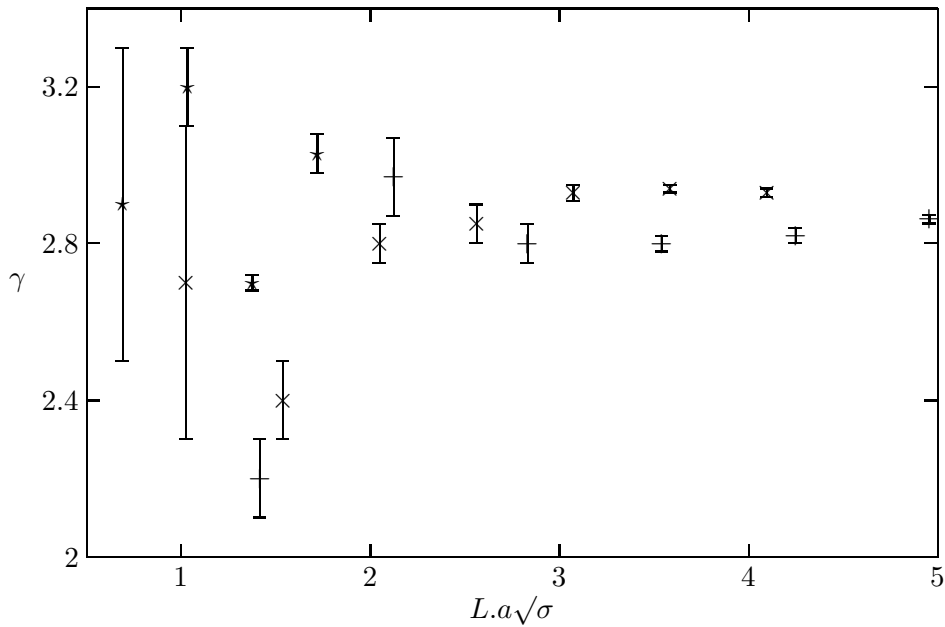
\begin	{figure}[t]
\begin	{flushright}
\leavevmode
\input{plot_gamm2}
\end	{flushright}

\vspace{\jumpback}
\caption{Scaling of the spectrum exponent $\gamma$. Key: $+$ -
$\beta=2.3$, $\times$ - $\beta=2.4$, $\star$ - $\beta=2.5$}
\label	{fig_plot_gamm}

\vspace{\jumpback}
\end 	{figure}

\noindent
1. The infinite volume spectrum is described exactly by the power law,
and loops of all sizes scale well. Finite volume effects distort this
above some multiple of the lattice size. In this picture, the
ambiguities encountered in resolving a cluster of monopole current
into a set of intersecting loops are of limited importance.

Small loops, however, appear to make no contribution to the
`asymptotic' string tension measured on large length scales
\cite{stack}.
It is unlikely that the correct string tension is given by
finite--size noise, which prejudices against this picture.

\noindent
2. The infinite volume spectrum is composed of two components. In
addition to the power law spectrum, there is also at least one large
current cluster of near constant current density which when placed in
a finite volume will give a current proportional to the volume. The
increase in the statistical errors for this portion comes from there
being no natural way of resolving this into closed loops.

This also poses questions. The large clusters, which apparently
generate the string tension, appear to recede with the lattice size
and have a current density which does not scale well. 
The string tension arises from disordering of the Wilson loop by
monopole structures. The monopole loop must have an extent that is
comparable to, or larger than, the Wilson loop if it is to disorder
it. Only the largest monopole loops will contribute to the `asymptotic'
string tension, but loops of intermediate size described by the power
law also would na\"{\i}vely be expected to contribute to the smaller
Creutz ratios. This contribution appears to be missing; the monopole
contribution to the inter-quark potential is extremely linear; $2
\times 2$ Creutz ratios differ from the asymptotic string tension by
only a few per cent.

One possible explanation is that the loops $l \leq \lbr$ are extremely
crumpled and their extent is not proportional to their length but less
(one-fourth power in the extreme case). Loops of length 50 would then form
objects that are only 3 units across. This is supported by
the rapid loss of all these loops under $U(1)$ cooling, without
significant decrease in the string tension
\cite{har_tep_prog}.

\section	{Other gauges and Gribov copies.}

\label		{sect_other_grib}

We may AP to gauges other than the MAG which diagonalise operators
such as; the Polyakov loop, the $U_{12}$ plaquette, $\sum U_{\mu\nu}
U_{\mu\nu}$, or `no gauge' where no diagonalisation is carried out.

The monopole spectrum is qualitatively the same; a power law breaking
off at long loops. The power law exponents are consistent with the
MAG. This is true even in `no gauge', where we expect no physics,
suggesting the value is of general, possibly combinatorial, origin.

There are more monopoles after projection to these gauges than to the
MAG. The values of $C$ are correspondingly higher and although scaling
well with {\em lattice} volume, scaling with {\em physical} length
scales is poor; $C$ is approximately constant across different
$\beta$, suggesting there is no contact with the long range physics.

Preferring a gauge where monopole structures behave physically will
cause us to favour the MAG, but may merely arise from the MAG operator
alone requiring iterative gauge fixing.

This iterative fixing allows Gribov copies. There is a negative
correlation between the value of the functional that is locally
maximised when gauge fixing and $c_p$. Scaling, however, remains good
and no motivation for the selection of Gribov copies is apparent from
this work so far.

\end{document}

%% file: plot_spec2.tex
\setlength{\unitlength}{0.240900pt}
\ifx\plotpoint\undefined\newsavebox{\plotpoint}\fi
\sbox{\plotpoint}{\rule[-0.200pt]{0.400pt}{0.400pt}}%
\begin{picture}(1500,900)(0,0)
\font\gnuplot=cmr10 at 12pt
\gnuplot
\sbox{\plotpoint}{\rule[-0.200pt]{0.400pt}{0.400pt}}%
\put(120.0,103.0){\rule[-0.200pt]{4.818pt}{0.400pt}}
\put(108,103){\makebox(0,0)[r]{{$-5$}}}
\put(1436.0,103.0){\rule[-0.200pt]{4.818pt}{0.400pt}}
\put(120.0,247.0){\rule[-0.200pt]{4.818pt}{0.400pt}}
\put(108,247){\makebox(0,0)[r]{{$-3$}}}
\put(1436.0,247.0){\rule[-0.200pt]{4.818pt}{0.400pt}}
\put(120.0,390.0){\rule[-0.200pt]{4.818pt}{0.400pt}}
\put(108,390){\makebox(0,0)[r]{{$-1$}}}
\put(1436.0,390.0){\rule[-0.200pt]{4.818pt}{0.400pt}}
\put(120.0,534.0){\rule[-0.200pt]{4.818pt}{0.400pt}}
\put(108,534){\makebox(0,0)[r]{{$1$}}}
\put(1436.0,534.0){\rule[-0.200pt]{4.818pt}{0.400pt}}
\put(120.0,678.0){\rule[-0.200pt]{4.818pt}{0.400pt}}
\put(108,678){\makebox(0,0)[r]{{$3$}}}
\put(1436.0,678.0){\rule[-0.200pt]{4.818pt}{0.400pt}}
\put(120.0,821.0){\rule[-0.200pt]{4.818pt}{0.400pt}}
\put(108,821){\makebox(0,0)[r]{{$5$}}}
\put(1436.0,821.0){\rule[-0.200pt]{4.818pt}{0.400pt}}
\put(120.0,31.0){\rule[-0.200pt]{0.400pt}{4.818pt}}
\put(120,19){\makebox(0,0){\shortstack{\\ \\ \\ {$1$}}}}
\put(120.0,873.0){\rule[-0.200pt]{0.400pt}{4.818pt}}
\put(417.0,31.0){\rule[-0.200pt]{0.400pt}{4.818pt}}
\put(417,19){\makebox(0,0){\shortstack{\\ \\ \\ {$2$}}}}
\put(417.0,873.0){\rule[-0.200pt]{0.400pt}{4.818pt}}
\put(714.0,31.0){\rule[-0.200pt]{0.400pt}{4.818pt}}
\put(714,19){\makebox(0,0){\shortstack{\\ \\ \\ {$3$}}}}
\put(714.0,873.0){\rule[-0.200pt]{0.400pt}{4.818pt}}
\put(1011.0,31.0){\rule[-0.200pt]{0.400pt}{4.818pt}}
\put(1011,19){\makebox(0,0){\shortstack{\\ \\ \\ {$4$}}}}
\put(1011.0,873.0){\rule[-0.200pt]{0.400pt}{4.818pt}}
\put(1308.0,31.0){\rule[-0.200pt]{0.400pt}{4.818pt}}
\put(1308,19){\makebox(0,0){\shortstack{\\ \\ \\ {$5$}}}}
\put(1308.0,873.0){\rule[-0.200pt]{0.400pt}{4.818pt}}
\put(120.0,31.0){\rule[-0.200pt]{321.842pt}{0.400pt}}
\put(1456.0,31.0){\rule[-0.200pt]{0.400pt}{207.656pt}}
\put(120.0,893.0){\rule[-0.200pt]{321.842pt}{0.400pt}}
\put(12,522){\makebox(0,0){{$\log N(l)$}}}
\put(788,-53){\makebox(0,0){{$\log l$}}}
\put(120.0,31.0){\rule[-0.200pt]{0.400pt}{207.656pt}}
\put(235,829){\circle{12}}
\put(355,761){\circle{12}}
\put(440,702){\circle{12}}
\put(507,654){\circle{12}}
\put(561,615){\circle{12}}
\put(607,582){\circle{12}}
\put(646,554){\circle{12}}
\put(681,530){\circle{12}}
\put(713,506){\circle{12}}
\put(741,487){\circle{12}}
\put(767,466){\circle{12}}
\put(790,453){\circle{12}}
\put(812,431){\circle{12}}
\put(833,416){\circle{12}}
\put(852,410){\circle{12}}
\put(870,403){\circle{12}}
\put(887,390){\circle{12}}
\put(903,375){\circle{12}}
\put(918,363){\circle{12}}
\put(933,351){\circle{12}}
\put(947,334){\circle{12}}
\put(960,341){\circle{12}}
\put(972,330){\circle{12}}
\put(985,323){\circle{12}}
\put(996,304){\circle{12}}
\put(1007,294){\circle{12}}
\put(1018,307){\circle{12}}
\put(1029,303){\circle{12}}
\put(1039,271){\circle{12}}
\put(1048,272){\circle{12}}
\put(1058,257){\circle{12}}
\put(1067,264){\circle{12}}
\put(1076,249){\circle{12}}
\put(1084,257){\circle{12}}
\put(1093,231){\circle{12}}
\put(1101,235){\circle{12}}
\put(1109,210){\circle{12}}
\put(1117,219){\circle{12}}
\put(1124,197){\circle{12}}
\put(1131,224){\circle{12}}
\put(1139,166){\circle{12}}
\put(1146,235){\circle{12}}
\put(1152,201){\circle{12}}
\put(1159,213){\circle{12}}
\put(1166,219){\circle{12}}
\put(1172,210){\circle{12}}
\put(1178,189){\circle{12}}
\put(1184,181){\circle{12}}
\put(1190,201){\circle{12}}
\put(1196,235){\circle{12}}
\put(1202,171){\circle{12}}
\put(1208,193){\circle{12}}
\put(1213,154){\circle{12}}
\put(1219,185){\circle{12}}
\put(1224,216){\circle{12}}
\put(1229,201){\circle{12}}
\put(1234,147){\circle{12}}
\put(1239,171){\circle{12}}
\put(1244,181){\circle{12}}
\put(1249,201){\circle{12}}
\put(1254,204){\circle{12}}
\put(1259,176){\circle{12}}
\put(1264,197){\circle{12}}
\put(1268,189){\circle{12}}
\put(1273,140){\circle{12}}
\put(1277,154){\circle{12}}
\put(1282,160){\circle{12}}
\put(1286,166){\circle{12}}
\put(1290,147){\circle{12}}
\put(1294,160){\circle{12}}
\put(1299,193){\circle{12}}
\put(1303,160){\circle{12}}
\put(1307,122){\circle{12}}
\put(1311,140){\circle{12}}
\put(1315,122){\circle{12}}
\put(1319,166){\circle{12}}
\put(1322,176){\circle{12}}
\put(1326,154){\circle{12}}
\put(1330,147){\circle{12}}
\put(1334,131){\circle{12}}
\put(1337,189){\circle{12}}
\put(1341,160){\circle{12}}
\put(1344,81){\circle{12}}
\put(1348,160){\circle{12}}
\put(1351,122){\circle{12}}
\put(1355,111){\circle{12}}
\put(1358,166){\circle{12}}
\put(1362,111){\circle{12}}
\put(1365,147){\circle{12}}
\put(1368,154){\circle{12}}
\put(1371,189){\circle{12}}
\put(1375,154){\circle{12}}
\put(1378,111){\circle{12}}
\put(1381,131){\circle{12}}
\put(1384,111){\circle{12}}
\put(1387,81){\circle{12}}
\put(1390,154){\circle{12}}
\put(1393,122){\circle{12}}
\put(235.0,828.0){\usebox{\plotpoint}}
\put(225.0,828.0){\rule[-0.200pt]{4.818pt}{0.400pt}}
\put(225.0,829.0){\rule[-0.200pt]{4.818pt}{0.400pt}}
\put(355.0,760.0){\usebox{\plotpoint}}
\put(345.0,760.0){\rule[-0.200pt]{4.818pt}{0.400pt}}
\put(345.0,761.0){\rule[-0.200pt]{4.818pt}{0.400pt}}
\put(440.0,702.0){\usebox{\plotpoint}}
\put(430.0,702.0){\rule[-0.200pt]{4.818pt}{0.400pt}}
\put(430.0,703.0){\rule[-0.200pt]{4.818pt}{0.400pt}}
\put(507.0,653.0){\usebox{\plotpoint}}
\put(497.0,653.0){\rule[-0.200pt]{4.818pt}{0.400pt}}
\put(497.0,654.0){\rule[-0.200pt]{4.818pt}{0.400pt}}
\put(561.0,614.0){\rule[-0.200pt]{0.400pt}{0.482pt}}
\put(551.0,614.0){\rule[-0.200pt]{4.818pt}{0.400pt}}
\put(551.0,616.0){\rule[-0.200pt]{4.818pt}{0.400pt}}
\put(607.0,581.0){\rule[-0.200pt]{0.400pt}{0.482pt}}
\put(597.0,581.0){\rule[-0.200pt]{4.818pt}{0.400pt}}
\put(597.0,583.0){\rule[-0.200pt]{4.818pt}{0.400pt}}
\put(646.0,553.0){\rule[-0.200pt]{0.400pt}{0.723pt}}
\put(636.0,553.0){\rule[-0.200pt]{4.818pt}{0.400pt}}
\put(636.0,556.0){\rule[-0.200pt]{4.818pt}{0.400pt}}
\put(681.0,528.0){\rule[-0.200pt]{0.400pt}{0.723pt}}
\put(671.0,528.0){\rule[-0.200pt]{4.818pt}{0.400pt}}
\put(671.0,531.0){\rule[-0.200pt]{4.818pt}{0.400pt}}
\put(713.0,504.0){\rule[-0.200pt]{0.400pt}{0.964pt}}
\put(703.0,504.0){\rule[-0.200pt]{4.818pt}{0.400pt}}
\put(703.0,508.0){\rule[-0.200pt]{4.818pt}{0.400pt}}
\put(741.0,485.0){\rule[-0.200pt]{0.400pt}{0.964pt}}
\put(731.0,485.0){\rule[-0.200pt]{4.818pt}{0.400pt}}
\put(731.0,489.0){\rule[-0.200pt]{4.818pt}{0.400pt}}
\put(767.0,463.0){\rule[-0.200pt]{0.400pt}{1.445pt}}
\put(757.0,463.0){\rule[-0.200pt]{4.818pt}{0.400pt}}
\put(757.0,469.0){\rule[-0.200pt]{4.818pt}{0.400pt}}
\put(790.0,450.0){\rule[-0.200pt]{0.400pt}{1.204pt}}
\put(780.0,450.0){\rule[-0.200pt]{4.818pt}{0.400pt}}
\put(780.0,455.0){\rule[-0.200pt]{4.818pt}{0.400pt}}
\put(812.0,428.0){\rule[-0.200pt]{0.400pt}{1.686pt}}
\put(802.0,428.0){\rule[-0.200pt]{4.818pt}{0.400pt}}
\put(802.0,435.0){\rule[-0.200pt]{4.818pt}{0.400pt}}
\put(833.0,412.0){\rule[-0.200pt]{0.400pt}{1.927pt}}
\put(823.0,412.0){\rule[-0.200pt]{4.818pt}{0.400pt}}
\put(823.0,420.0){\rule[-0.200pt]{4.818pt}{0.400pt}}
\put(852.0,406.0){\rule[-0.200pt]{0.400pt}{1.927pt}}
\put(842.0,406.0){\rule[-0.200pt]{4.818pt}{0.400pt}}
\put(842.0,414.0){\rule[-0.200pt]{4.818pt}{0.400pt}}
\put(870.0,399.0){\rule[-0.200pt]{0.400pt}{2.168pt}}
\put(860.0,399.0){\rule[-0.200pt]{4.818pt}{0.400pt}}
\put(860.0,408.0){\rule[-0.200pt]{4.818pt}{0.400pt}}
\put(887.0,386.0){\rule[-0.200pt]{0.400pt}{2.168pt}}
\put(877.0,386.0){\rule[-0.200pt]{4.818pt}{0.400pt}}
\put(877.0,395.0){\rule[-0.200pt]{4.818pt}{0.400pt}}
\put(903.0,371.0){\rule[-0.200pt]{0.400pt}{2.168pt}}
\put(893.0,371.0){\rule[-0.200pt]{4.818pt}{0.400pt}}
\put(893.0,380.0){\rule[-0.200pt]{4.818pt}{0.400pt}}
\put(918.0,357.0){\rule[-0.200pt]{0.400pt}{2.650pt}}
\put(908.0,357.0){\rule[-0.200pt]{4.818pt}{0.400pt}}
\put(908.0,368.0){\rule[-0.200pt]{4.818pt}{0.400pt}}
\put(933.0,346.0){\rule[-0.200pt]{0.400pt}{2.409pt}}
\put(923.0,346.0){\rule[-0.200pt]{4.818pt}{0.400pt}}
\put(923.0,356.0){\rule[-0.200pt]{4.818pt}{0.400pt}}
\put(947.0,327.0){\rule[-0.200pt]{0.400pt}{3.132pt}}
\put(937.0,327.0){\rule[-0.200pt]{4.818pt}{0.400pt}}
\put(937.0,340.0){\rule[-0.200pt]{4.818pt}{0.400pt}}
\put(960.0,336.0){\rule[-0.200pt]{0.400pt}{2.650pt}}
\put(950.0,336.0){\rule[-0.200pt]{4.818pt}{0.400pt}}
\put(950.0,347.0){\rule[-0.200pt]{4.818pt}{0.400pt}}
\put(972.0,324.0){\rule[-0.200pt]{0.400pt}{2.891pt}}
\put(962.0,324.0){\rule[-0.200pt]{4.818pt}{0.400pt}}
\put(962.0,336.0){\rule[-0.200pt]{4.818pt}{0.400pt}}
\put(985.0,316.0){\rule[-0.200pt]{0.400pt}{3.132pt}}
\put(975.0,316.0){\rule[-0.200pt]{4.818pt}{0.400pt}}
\put(975.0,329.0){\rule[-0.200pt]{4.818pt}{0.400pt}}
\put(996.0,297.0){\rule[-0.200pt]{0.400pt}{3.613pt}}
\put(986.0,297.0){\rule[-0.200pt]{4.818pt}{0.400pt}}
\put(986.0,312.0){\rule[-0.200pt]{4.818pt}{0.400pt}}
\put(1007.0,286.0){\rule[-0.200pt]{0.400pt}{3.613pt}}
\put(997.0,286.0){\rule[-0.200pt]{4.818pt}{0.400pt}}
\put(997.0,301.0){\rule[-0.200pt]{4.818pt}{0.400pt}}
\put(1018.0,300.0){\rule[-0.200pt]{0.400pt}{3.613pt}}
\put(1008.0,300.0){\rule[-0.200pt]{4.818pt}{0.400pt}}
\put(1008.0,315.0){\rule[-0.200pt]{4.818pt}{0.400pt}}
\put(1029.0,294.0){\rule[-0.200pt]{0.400pt}{4.095pt}}
\put(1019.0,294.0){\rule[-0.200pt]{4.818pt}{0.400pt}}
\put(1019.0,311.0){\rule[-0.200pt]{4.818pt}{0.400pt}}
\put(1039.0,262.0){\rule[-0.200pt]{0.400pt}{4.336pt}}
\put(1029.0,262.0){\rule[-0.200pt]{4.818pt}{0.400pt}}
\put(1029.0,280.0){\rule[-0.200pt]{4.818pt}{0.400pt}}
\put(1048.0,263.0){\rule[-0.200pt]{0.400pt}{4.577pt}}
\put(1038.0,263.0){\rule[-0.200pt]{4.818pt}{0.400pt}}
\put(1038.0,282.0){\rule[-0.200pt]{4.818pt}{0.400pt}}
\put(1058.0,247.0){\rule[-0.200pt]{0.400pt}{4.818pt}}
\put(1048.0,247.0){\rule[-0.200pt]{4.818pt}{0.400pt}}
\put(1048.0,267.0){\rule[-0.200pt]{4.818pt}{0.400pt}}
\put(1067.0,254.0){\rule[-0.200pt]{0.400pt}{5.059pt}}
\put(1057.0,254.0){\rule[-0.200pt]{4.818pt}{0.400pt}}
\put(1057.0,275.0){\rule[-0.200pt]{4.818pt}{0.400pt}}
\put(1076.0,237.0){\rule[-0.200pt]{0.400pt}{5.541pt}}
\put(1066.0,237.0){\rule[-0.200pt]{4.818pt}{0.400pt}}
\put(1066.0,260.0){\rule[-0.200pt]{4.818pt}{0.400pt}}
\put(1084.0,246.0){\rule[-0.200pt]{0.400pt}{5.300pt}}
\put(1074.0,246.0){\rule[-0.200pt]{4.818pt}{0.400pt}}
\put(1074.0,268.0){\rule[-0.200pt]{4.818pt}{0.400pt}}
\put(1093.0,219.0){\rule[-0.200pt]{0.400pt}{5.782pt}}
\put(1083.0,219.0){\rule[-0.200pt]{4.818pt}{0.400pt}}
\put(1083.0,243.0){\rule[-0.200pt]{4.818pt}{0.400pt}}
\put(1101.0,222.0){\rule[-0.200pt]{0.400pt}{6.504pt}}
\put(1091.0,222.0){\rule[-0.200pt]{4.818pt}{0.400pt}}
\put(1091.0,249.0){\rule[-0.200pt]{4.818pt}{0.400pt}}
\put(1109.0,197.0){\rule[-0.200pt]{0.400pt}{6.263pt}}
\put(1099.0,197.0){\rule[-0.200pt]{4.818pt}{0.400pt}}
\put(1099.0,223.0){\rule[-0.200pt]{4.818pt}{0.400pt}}
\put(1117.0,205.0){\rule[-0.200pt]{0.400pt}{6.504pt}}
\put(1107.0,205.0){\rule[-0.200pt]{4.818pt}{0.400pt}}
\put(1107.0,232.0){\rule[-0.200pt]{4.818pt}{0.400pt}}
\put(1124.0,180.0){\rule[-0.200pt]{0.400pt}{8.191pt}}
\put(1114.0,180.0){\rule[-0.200pt]{4.818pt}{0.400pt}}
\put(1114.0,214.0){\rule[-0.200pt]{4.818pt}{0.400pt}}
\put(1131.0,208.0){\rule[-0.200pt]{0.400pt}{7.468pt}}
\put(1121.0,208.0){\rule[-0.200pt]{4.818pt}{0.400pt}}
\put(1121.0,239.0){\rule[-0.200pt]{4.818pt}{0.400pt}}
\put(1139.0,146.0){\rule[-0.200pt]{0.400pt}{9.636pt}}
\put(1129.0,146.0){\rule[-0.200pt]{4.818pt}{0.400pt}}
\put(1129.0,186.0){\rule[-0.200pt]{4.818pt}{0.400pt}}
\put(1146.0,224.0){\rule[-0.200pt]{0.400pt}{5.541pt}}
\put(1136.0,224.0){\rule[-0.200pt]{4.818pt}{0.400pt}}
\put(1136.0,247.0){\rule[-0.200pt]{4.818pt}{0.400pt}}
\put(1152.0,183.0){\rule[-0.200pt]{0.400pt}{8.431pt}}
\put(1142.0,183.0){\rule[-0.200pt]{4.818pt}{0.400pt}}
\put(1142.0,218.0){\rule[-0.200pt]{4.818pt}{0.400pt}}
\put(1159.0,196.0){\rule[-0.200pt]{0.400pt}{8.191pt}}
\put(1149.0,196.0){\rule[-0.200pt]{4.818pt}{0.400pt}}
\put(1149.0,230.0){\rule[-0.200pt]{4.818pt}{0.400pt}}
\put(1166.0,204.0){\rule[-0.200pt]{0.400pt}{6.986pt}}
\put(1156.0,204.0){\rule[-0.200pt]{4.818pt}{0.400pt}}
\put(1156.0,233.0){\rule[-0.200pt]{4.818pt}{0.400pt}}
\put(1172.0,193.0){\rule[-0.200pt]{0.400pt}{8.191pt}}
\put(1162.0,193.0){\rule[-0.200pt]{4.818pt}{0.400pt}}
\put(1162.0,227.0){\rule[-0.200pt]{4.818pt}{0.400pt}}
\put(1178.0,172.0){\rule[-0.200pt]{0.400pt}{8.431pt}}
\put(1168.0,172.0){\rule[-0.200pt]{4.818pt}{0.400pt}}
\put(1168.0,207.0){\rule[-0.200pt]{4.818pt}{0.400pt}}
\put(1184.0,161.0){\rule[-0.200pt]{0.400pt}{9.636pt}}
\put(1174.0,161.0){\rule[-0.200pt]{4.818pt}{0.400pt}}
\put(1174.0,201.0){\rule[-0.200pt]{4.818pt}{0.400pt}}
\put(1190.0,181.0){\rule[-0.200pt]{0.400pt}{9.395pt}}
\put(1180.0,181.0){\rule[-0.200pt]{4.818pt}{0.400pt}}
\put(1180.0,220.0){\rule[-0.200pt]{4.818pt}{0.400pt}}
\put(1196.0,223.0){\rule[-0.200pt]{0.400pt}{5.782pt}}
\put(1186.0,223.0){\rule[-0.200pt]{4.818pt}{0.400pt}}
\put(1186.0,247.0){\rule[-0.200pt]{4.818pt}{0.400pt}}
\put(1202.0,152.0){\rule[-0.200pt]{0.400pt}{9.154pt}}
\put(1192.0,152.0){\rule[-0.200pt]{4.818pt}{0.400pt}}
\put(1192.0,190.0){\rule[-0.200pt]{4.818pt}{0.400pt}}
\put(1208.0,180.0){\rule[-0.200pt]{0.400pt}{6.504pt}}
\put(1198.0,180.0){\rule[-0.200pt]{4.818pt}{0.400pt}}
\put(1198.0,207.0){\rule[-0.200pt]{4.818pt}{0.400pt}}
\put(1213.0,134.0){\rule[-0.200pt]{0.400pt}{9.636pt}}
\put(1203.0,134.0){\rule[-0.200pt]{4.818pt}{0.400pt}}
\put(1203.0,174.0){\rule[-0.200pt]{4.818pt}{0.400pt}}
\put(1219.0,167.0){\rule[-0.200pt]{0.400pt}{8.913pt}}
\put(1209.0,167.0){\rule[-0.200pt]{4.818pt}{0.400pt}}
\put(1209.0,204.0){\rule[-0.200pt]{4.818pt}{0.400pt}}
\put(1224.0,200.0){\rule[-0.200pt]{0.400pt}{7.468pt}}
\put(1214.0,200.0){\rule[-0.200pt]{4.818pt}{0.400pt}}
\put(1214.0,231.0){\rule[-0.200pt]{4.818pt}{0.400pt}}
\put(1229.0,180.0){\rule[-0.200pt]{0.400pt}{9.877pt}}
\put(1219.0,180.0){\rule[-0.200pt]{4.818pt}{0.400pt}}
\put(1219.0,221.0){\rule[-0.200pt]{4.818pt}{0.400pt}}
\put(1234.0,123.0){\rule[-0.200pt]{0.400pt}{11.804pt}}
\put(1224.0,123.0){\rule[-0.200pt]{4.818pt}{0.400pt}}
\put(1224.0,172.0){\rule[-0.200pt]{4.818pt}{0.400pt}}
\put(1239.0,151.0){\rule[-0.200pt]{0.400pt}{9.877pt}}
\put(1229.0,151.0){\rule[-0.200pt]{4.818pt}{0.400pt}}
\put(1229.0,192.0){\rule[-0.200pt]{4.818pt}{0.400pt}}
\put(1244.0,164.0){\rule[-0.200pt]{0.400pt}{8.191pt}}
\put(1234.0,164.0){\rule[-0.200pt]{4.818pt}{0.400pt}}
\put(1234.0,198.0){\rule[-0.200pt]{4.818pt}{0.400pt}}
\put(1249.0,184.0){\rule[-0.200pt]{0.400pt}{8.191pt}}
\put(1239.0,184.0){\rule[-0.200pt]{4.818pt}{0.400pt}}
\put(1239.0,218.0){\rule[-0.200pt]{4.818pt}{0.400pt}}
\put(1254.0,188.0){\rule[-0.200pt]{0.400pt}{7.709pt}}
\put(1244.0,188.0){\rule[-0.200pt]{4.818pt}{0.400pt}}
\put(1244.0,220.0){\rule[-0.200pt]{4.818pt}{0.400pt}}
\put(1259.0,157.0){\rule[-0.200pt]{0.400pt}{9.395pt}}
\put(1249.0,157.0){\rule[-0.200pt]{4.818pt}{0.400pt}}
\put(1249.0,196.0){\rule[-0.200pt]{4.818pt}{0.400pt}}
\put(1264.0,180.0){\rule[-0.200pt]{0.400pt}{8.191pt}}
\put(1254.0,180.0){\rule[-0.200pt]{4.818pt}{0.400pt}}
\put(1254.0,214.0){\rule[-0.200pt]{4.818pt}{0.400pt}}
\put(1268.0,174.0){\rule[-0.200pt]{0.400pt}{7.468pt}}
\put(1258.0,174.0){\rule[-0.200pt]{4.818pt}{0.400pt}}
\put(1258.0,205.0){\rule[-0.200pt]{4.818pt}{0.400pt}}
\put(1273.0,113.0){\rule[-0.200pt]{0.400pt}{13.009pt}}
\put(1263.0,113.0){\rule[-0.200pt]{4.818pt}{0.400pt}}
\put(1263.0,167.0){\rule[-0.200pt]{4.818pt}{0.400pt}}
\put(1277.0,126.0){\rule[-0.200pt]{0.400pt}{13.490pt}}
\put(1267.0,126.0){\rule[-0.200pt]{4.818pt}{0.400pt}}
\put(1267.0,182.0){\rule[-0.200pt]{4.818pt}{0.400pt}}
\put(1282.0,139.0){\rule[-0.200pt]{0.400pt}{10.359pt}}
\put(1272.0,139.0){\rule[-0.200pt]{4.818pt}{0.400pt}}
\put(1272.0,182.0){\rule[-0.200pt]{4.818pt}{0.400pt}}
\put(1286.0,146.0){\rule[-0.200pt]{0.400pt}{9.636pt}}
\put(1276.0,146.0){\rule[-0.200pt]{4.818pt}{0.400pt}}
\put(1276.0,186.0){\rule[-0.200pt]{4.818pt}{0.400pt}}
\put(1290.0,123.0){\rule[-0.200pt]{0.400pt}{11.804pt}}
\put(1280.0,123.0){\rule[-0.200pt]{4.818pt}{0.400pt}}
\put(1280.0,172.0){\rule[-0.200pt]{4.818pt}{0.400pt}}
\put(1294.0,133.0){\rule[-0.200pt]{0.400pt}{13.249pt}}
\put(1284.0,133.0){\rule[-0.200pt]{4.818pt}{0.400pt}}
\put(1284.0,188.0){\rule[-0.200pt]{4.818pt}{0.400pt}}
\put(1299.0,174.0){\rule[-0.200pt]{0.400pt}{9.154pt}}
\put(1289.0,174.0){\rule[-0.200pt]{4.818pt}{0.400pt}}
\put(1289.0,212.0){\rule[-0.200pt]{4.818pt}{0.400pt}}
\put(1303.0,139.0){\rule[-0.200pt]{0.400pt}{10.359pt}}
\put(1293.0,139.0){\rule[-0.200pt]{4.818pt}{0.400pt}}
\put(1293.0,182.0){\rule[-0.200pt]{4.818pt}{0.400pt}}
\put(1307.0,94.0){\rule[-0.200pt]{0.400pt}{13.249pt}}
\put(1297.0,94.0){\rule[-0.200pt]{4.818pt}{0.400pt}}
\put(1297.0,149.0){\rule[-0.200pt]{4.818pt}{0.400pt}}
\put(1311.0,110.0){\rule[-0.200pt]{0.400pt}{14.454pt}}
\put(1301.0,110.0){\rule[-0.200pt]{4.818pt}{0.400pt}}
\put(1301.0,170.0){\rule[-0.200pt]{4.818pt}{0.400pt}}
\put(1315.0,94.0){\rule[-0.200pt]{0.400pt}{13.249pt}}
\put(1305.0,94.0){\rule[-0.200pt]{4.818pt}{0.400pt}}
\put(1305.0,149.0){\rule[-0.200pt]{4.818pt}{0.400pt}}
\put(1319.0,142.0){\rule[-0.200pt]{0.400pt}{11.563pt}}
\put(1309.0,142.0){\rule[-0.200pt]{4.818pt}{0.400pt}}
\put(1309.0,190.0){\rule[-0.200pt]{4.818pt}{0.400pt}}
\put(1322.0,156.0){\rule[-0.200pt]{0.400pt}{9.877pt}}
\put(1312.0,156.0){\rule[-0.200pt]{4.818pt}{0.400pt}}
\put(1312.0,197.0){\rule[-0.200pt]{4.818pt}{0.400pt}}
\put(1326.0,131.0){\rule[-0.200pt]{0.400pt}{11.081pt}}
\put(1316.0,131.0){\rule[-0.200pt]{4.818pt}{0.400pt}}
\put(1316.0,177.0){\rule[-0.200pt]{4.818pt}{0.400pt}}
\put(1330.0,123.0){\rule[-0.200pt]{0.400pt}{11.804pt}}
\put(1320.0,123.0){\rule[-0.200pt]{4.818pt}{0.400pt}}
\put(1320.0,172.0){\rule[-0.200pt]{4.818pt}{0.400pt}}
\put(1334.0,106.0){\rule[-0.200pt]{0.400pt}{12.045pt}}
\put(1324.0,106.0){\rule[-0.200pt]{4.818pt}{0.400pt}}
\put(1324.0,156.0){\rule[-0.200pt]{4.818pt}{0.400pt}}
\put(1337.0,175.0){\rule[-0.200pt]{0.400pt}{6.986pt}}
\put(1327.0,175.0){\rule[-0.200pt]{4.818pt}{0.400pt}}
\put(1327.0,204.0){\rule[-0.200pt]{4.818pt}{0.400pt}}
\put(1341.0,139.0){\rule[-0.200pt]{0.400pt}{10.359pt}}
\put(1331.0,139.0){\rule[-0.200pt]{4.818pt}{0.400pt}}
\put(1331.0,182.0){\rule[-0.200pt]{4.818pt}{0.400pt}}
\put(1344.0,41.0){\rule[-0.200pt]{0.400pt}{19.272pt}}
\put(1334.0,41.0){\rule[-0.200pt]{4.818pt}{0.400pt}}
\put(1334.0,121.0){\rule[-0.200pt]{4.818pt}{0.400pt}}
\put(1348.0,137.0){\rule[-0.200pt]{0.400pt}{11.322pt}}
\put(1338.0,137.0){\rule[-0.200pt]{4.818pt}{0.400pt}}
\put(1338.0,184.0){\rule[-0.200pt]{4.818pt}{0.400pt}}
\put(1351.0,94.0){\rule[-0.200pt]{0.400pt}{13.249pt}}
\put(1341.0,94.0){\rule[-0.200pt]{4.818pt}{0.400pt}}
\put(1341.0,149.0){\rule[-0.200pt]{4.818pt}{0.400pt}}
\put(1355.0,80.0){\rule[-0.200pt]{0.400pt}{14.695pt}}
\put(1345.0,80.0){\rule[-0.200pt]{4.818pt}{0.400pt}}
\put(1345.0,141.0){\rule[-0.200pt]{4.818pt}{0.400pt}}
\put(1358.0,144.0){\rule[-0.200pt]{0.400pt}{10.600pt}}
\put(1348.0,144.0){\rule[-0.200pt]{4.818pt}{0.400pt}}
\put(1348.0,188.0){\rule[-0.200pt]{4.818pt}{0.400pt}}
\put(1362.0,73.0){\rule[-0.200pt]{0.400pt}{18.067pt}}
\put(1352.0,73.0){\rule[-0.200pt]{4.818pt}{0.400pt}}
\put(1352.0,148.0){\rule[-0.200pt]{4.818pt}{0.400pt}}
\put(1365.0,126.0){\rule[-0.200pt]{0.400pt}{10.359pt}}
\put(1355.0,126.0){\rule[-0.200pt]{4.818pt}{0.400pt}}
\put(1355.0,169.0){\rule[-0.200pt]{4.818pt}{0.400pt}}
\put(1368.0,129.0){\rule[-0.200pt]{0.400pt}{12.045pt}}
\put(1358.0,129.0){\rule[-0.200pt]{4.818pt}{0.400pt}}
\put(1358.0,179.0){\rule[-0.200pt]{4.818pt}{0.400pt}}
\put(1371.0,172.0){\rule[-0.200pt]{0.400pt}{8.431pt}}
\put(1361.0,172.0){\rule[-0.200pt]{4.818pt}{0.400pt}}
\put(1361.0,207.0){\rule[-0.200pt]{4.818pt}{0.400pt}}
\put(1375.0,127.0){\rule[-0.200pt]{0.400pt}{13.249pt}}
\put(1365.0,127.0){\rule[-0.200pt]{4.818pt}{0.400pt}}
\put(1365.0,182.0){\rule[-0.200pt]{4.818pt}{0.400pt}}
\put(1378.0,80.0){\rule[-0.200pt]{0.400pt}{14.695pt}}
\put(1368.0,80.0){\rule[-0.200pt]{4.818pt}{0.400pt}}
\put(1368.0,141.0){\rule[-0.200pt]{4.818pt}{0.400pt}}
\put(1381.0,106.0){\rule[-0.200pt]{0.400pt}{12.045pt}}
\put(1371.0,106.0){\rule[-0.200pt]{4.818pt}{0.400pt}}
\put(1371.0,156.0){\rule[-0.200pt]{4.818pt}{0.400pt}}
\put(1384.0,80.0){\rule[-0.200pt]{0.400pt}{14.695pt}}
\put(1374.0,80.0){\rule[-0.200pt]{4.818pt}{0.400pt}}
\put(1374.0,141.0){\rule[-0.200pt]{4.818pt}{0.400pt}}
\put(1387.0,41.0){\rule[-0.200pt]{0.400pt}{19.272pt}}
\put(1377.0,41.0){\rule[-0.200pt]{4.818pt}{0.400pt}}
\put(1377.0,121.0){\rule[-0.200pt]{4.818pt}{0.400pt}}
\put(1390.0,131.0){\rule[-0.200pt]{0.400pt}{11.081pt}}
\put(1380.0,131.0){\rule[-0.200pt]{4.818pt}{0.400pt}}
\put(1380.0,177.0){\rule[-0.200pt]{4.818pt}{0.400pt}}
\put(1393.0,77.0){\rule[-0.200pt]{0.400pt}{21.440pt}}
\put(1383.0,77.0){\rule[-0.200pt]{4.818pt}{0.400pt}}
\put(1383.0,166.0){\rule[-0.200pt]{4.818pt}{0.400pt}}
\end{picture}

%% file: plot_gamm2.tex
\setlength{\unitlength}{0.240900pt}
\ifx\plotpoint\undefined\newsavebox{\plotpoint}\fi
\sbox{\plotpoint}{\rule[-0.200pt]{0.400pt}{0.400pt}}%
\begin{picture}(1500,900)(0,0)
\font\gnuplot=cmr10 at 12pt
\gnuplot
\sbox{\plotpoint}{\rule[-0.200pt]{0.400pt}{0.400pt}}%
\put(120.0,31.0){\rule[-0.200pt]{4.818pt}{0.400pt}}
\put(108,31){\makebox(0,0)[r]{{$2$}}}
\put(1436.0,31.0){\rule[-0.200pt]{4.818pt}{0.400pt}}
\put(120.0,277.0){\rule[-0.200pt]{4.818pt}{0.400pt}}
\put(108,277){\makebox(0,0)[r]{{$2.4$}}}
\put(1436.0,277.0){\rule[-0.200pt]{4.818pt}{0.400pt}}
\put(120.0,524.0){\rule[-0.200pt]{4.818pt}{0.400pt}}
\put(108,524){\makebox(0,0)[r]{{$2.8$}}}
\put(1436.0,524.0){\rule[-0.200pt]{4.818pt}{0.400pt}}
\put(120.0,770.0){\rule[-0.200pt]{4.818pt}{0.400pt}}
\put(108,770){\makebox(0,0)[r]{{$3.2$}}}
\put(1436.0,770.0){\rule[-0.200pt]{4.818pt}{0.400pt}}
\put(268.0,31.0){\rule[-0.200pt]{0.400pt}{4.818pt}}
\put(268,19){\makebox(0,0){\shortstack{\\ \\ \\ {$1$}}}}
\put(268.0,873.0){\rule[-0.200pt]{0.400pt}{4.818pt}}
\put(565.0,31.0){\rule[-0.200pt]{0.400pt}{4.818pt}}
\put(565,19){\makebox(0,0){\shortstack{\\ \\ \\ {$2$}}}}
\put(565.0,873.0){\rule[-0.200pt]{0.400pt}{4.818pt}}
\put(862.0,31.0){\rule[-0.200pt]{0.400pt}{4.818pt}}
\put(862,19){\makebox(0,0){\shortstack{\\ \\ \\ {$3$}}}}
\put(862.0,873.0){\rule[-0.200pt]{0.400pt}{4.818pt}}
\put(1159.0,31.0){\rule[-0.200pt]{0.400pt}{4.818pt}}
\put(1159,19){\makebox(0,0){\shortstack{\\ \\ \\ {$4$}}}}
\put(1159.0,873.0){\rule[-0.200pt]{0.400pt}{4.818pt}}
\put(1456.0,31.0){\rule[-0.200pt]{0.400pt}{4.818pt}}
\put(1456,19){\makebox(0,0){\shortstack{\\ \\ \\ {$5$}}}}
\put(1456.0,873.0){\rule[-0.200pt]{0.400pt}{4.818pt}}
\put(120.0,31.0){\rule[-0.200pt]{321.842pt}{0.400pt}}
\put(1456.0,31.0){\rule[-0.200pt]{0.400pt}{207.656pt}}
\put(120.0,893.0){\rule[-0.200pt]{321.842pt}{0.400pt}}
\put(12,462){\makebox(0,0){{$\gamma$}}}
\put(788,-53){\makebox(0,0){{$L.a\surd\sigma$}}}
\put(120.0,31.0){\rule[-0.200pt]{0.400pt}{207.656pt}}
\put(392,154){\makebox(0,0){$+$}}
\put(602,628){\makebox(0,0){$+$}}
\put(812,524){\makebox(0,0){$+$}}
\put(1023,524){\makebox(0,0){$+$}}
\put(1233,536){\makebox(0,0){$+$}}
\put(1443,562){\makebox(0,0){$+$}}
\put(392.0,93.0){\rule[-0.200pt]{0.400pt}{29.631pt}}
\put(382.0,93.0){\rule[-0.200pt]{4.818pt}{0.400pt}}
\put(382.0,216.0){\rule[-0.200pt]{4.818pt}{0.400pt}}
\put(602.0,567.0){\rule[-0.200pt]{0.400pt}{29.631pt}}
\put(592.0,567.0){\rule[-0.200pt]{4.818pt}{0.400pt}}
\put(592.0,690.0){\rule[-0.200pt]{4.818pt}{0.400pt}}
\put(812.0,493.0){\rule[-0.200pt]{0.400pt}{14.695pt}}
\put(802.0,493.0){\rule[-0.200pt]{4.818pt}{0.400pt}}
\put(802.0,554.0){\rule[-0.200pt]{4.818pt}{0.400pt}}
\put(1023.0,511.0){\rule[-0.200pt]{0.400pt}{6.022pt}}
\put(1013.0,511.0){\rule[-0.200pt]{4.818pt}{0.400pt}}
\put(1013.0,536.0){\rule[-0.200pt]{4.818pt}{0.400pt}}
\put(1233.0,524.0){\rule[-0.200pt]{0.400pt}{5.782pt}}
\put(1223.0,524.0){\rule[-0.200pt]{4.818pt}{0.400pt}}
\put(1223.0,548.0){\rule[-0.200pt]{4.818pt}{0.400pt}}
\put(1443.0,555.0){\rule[-0.200pt]{0.400pt}{3.373pt}}
\put(1433.0,555.0){\rule[-0.200pt]{4.818pt}{0.400pt}}
\put(1433.0,569.0){\rule[-0.200pt]{4.818pt}{0.400pt}}
\put(276,462){\makebox(0,0){$\times$}}
\put(428,277){\makebox(0,0){$\times$}}
\put(580,524){\makebox(0,0){$\times$}}
\put(732,554){\makebox(0,0){$\times$}}
\put(884,604){\makebox(0,0){$\times$}}
\put(1036,610){\makebox(0,0){$\times$}}
\put(1188,604){\makebox(0,0){$\times$}}
\put(276.0,216.0){\rule[-0.200pt]{0.400pt}{118.523pt}}
\put(266.0,216.0){\rule[-0.200pt]{4.818pt}{0.400pt}}
\put(266.0,708.0){\rule[-0.200pt]{4.818pt}{0.400pt}}
\put(428.0,216.0){\rule[-0.200pt]{0.400pt}{29.631pt}}
\put(418.0,216.0){\rule[-0.200pt]{4.818pt}{0.400pt}}
\put(418.0,339.0){\rule[-0.200pt]{4.818pt}{0.400pt}}
\put(580.0,493.0){\rule[-0.200pt]{0.400pt}{14.695pt}}
\put(570.0,493.0){\rule[-0.200pt]{4.818pt}{0.400pt}}
\put(570.0,554.0){\rule[-0.200pt]{4.818pt}{0.400pt}}
\put(732.0,524.0){\rule[-0.200pt]{0.400pt}{14.695pt}}
\put(722.0,524.0){\rule[-0.200pt]{4.818pt}{0.400pt}}
\put(722.0,585.0){\rule[-0.200pt]{4.818pt}{0.400pt}}
\put(884.0,591.0){\rule[-0.200pt]{0.400pt}{6.022pt}}
\put(874.0,591.0){\rule[-0.200pt]{4.818pt}{0.400pt}}
\put(874.0,616.0){\rule[-0.200pt]{4.818pt}{0.400pt}}
\put(1036.0,604.0){\rule[-0.200pt]{0.400pt}{2.891pt}}
\put(1026.0,604.0){\rule[-0.200pt]{4.818pt}{0.400pt}}
\put(1026.0,616.0){\rule[-0.200pt]{4.818pt}{0.400pt}}
\put(1188.0,597.0){\rule[-0.200pt]{0.400pt}{3.132pt}}
\put(1178.0,597.0){\rule[-0.200pt]{4.818pt}{0.400pt}}
\put(1178.0,610.0){\rule[-0.200pt]{4.818pt}{0.400pt}}
\put(176,585){\makebox(0,0){$\star$}}
\put(278,770){\makebox(0,0){$\star$}}
\put(380,462){\makebox(0,0){$\star$}}
\put(482,665){\makebox(0,0){$\star$}}
\put(176.0,339.0){\rule[-0.200pt]{0.400pt}{118.523pt}}
\put(166.0,339.0){\rule[-0.200pt]{4.818pt}{0.400pt}}
\put(166.0,831.0){\rule[-0.200pt]{4.818pt}{0.400pt}}
\put(278.0,708.0){\rule[-0.200pt]{0.400pt}{29.631pt}}
\put(268.0,708.0){\rule[-0.200pt]{4.818pt}{0.400pt}}
\put(268.0,831.0){\rule[-0.200pt]{4.818pt}{0.400pt}}
\put(380.0,450.0){\rule[-0.200pt]{0.400pt}{5.782pt}}
\put(370.0,450.0){\rule[-0.200pt]{4.818pt}{0.400pt}}
\put(370.0,474.0){\rule[-0.200pt]{4.818pt}{0.400pt}}
\put(482.0,634.0){\rule[-0.200pt]{0.400pt}{14.936pt}}
\put(472.0,634.0){\rule[-0.200pt]{4.818pt}{0.400pt}}
\put(472.0,696.0){\rule[-0.200pt]{4.818pt}{0.400pt}}
\end{picture}

%% file: plot_cp2.tex
\setlength{\unitlength}{0.240900pt}
\ifx\plotpoint\undefined\newsavebox{\plotpoint}\fi
\begin{picture}(1500,900)(0,0)
\font\gnuplot=cmr10 at 12pt
\gnuplot
\sbox{\plotpoint}{\rule[-0.200pt]{0.400pt}{0.400pt}}%
\put(120.0,31.0){\rule[-0.200pt]{321.842pt}{0.400pt}}
\put(120.0,31.0){\rule[-0.200pt]{4.818pt}{0.400pt}}
\put(108,31){\makebox(0,0)[r]{{$0$}}}
\put(1436.0,31.0){\rule[-0.200pt]{4.818pt}{0.400pt}}
\put(120.0,318.0){\rule[-0.200pt]{4.818pt}{0.400pt}}
\put(108,318){\makebox(0,0)[r]{{$1$}}}
\put(1436.0,318.0){\rule[-0.200pt]{4.818pt}{0.400pt}}
\put(120.0,606.0){\rule[-0.200pt]{4.818pt}{0.400pt}}
\put(108,606){\makebox(0,0)[r]{{$2$}}}
\put(1436.0,606.0){\rule[-0.200pt]{4.818pt}{0.400pt}}
\put(120.0,893.0){\rule[-0.200pt]{4.818pt}{0.400pt}}
\put(108,893){\makebox(0,0)[r]{{$3$}}}
\put(1436.0,893.0){\rule[-0.200pt]{4.818pt}{0.400pt}}
\put(268.0,31.0){\rule[-0.200pt]{0.400pt}{4.818pt}}
\put(268,19){\makebox(0,0){\shortstack{\\ \\ \\ {$1$}}}}
\put(268.0,873.0){\rule[-0.200pt]{0.400pt}{4.818pt}}
\put(565.0,31.0){\rule[-0.200pt]{0.400pt}{4.818pt}}
\put(565,19){\makebox(0,0){\shortstack{\\ \\ \\ {$2$}}}}
\put(565.0,873.0){\rule[-0.200pt]{0.400pt}{4.818pt}}
\put(862.0,31.0){\rule[-0.200pt]{0.400pt}{4.818pt}}
\put(862,19){\makebox(0,0){\shortstack{\\ \\ \\ {$3$}}}}
\put(862.0,873.0){\rule[-0.200pt]{0.400pt}{4.818pt}}
\put(1159.0,31.0){\rule[-0.200pt]{0.400pt}{4.818pt}}
\put(1159,19){\makebox(0,0){\shortstack{\\ \\ \\ {$4$}}}}
\put(1159.0,873.0){\rule[-0.200pt]{0.400pt}{4.818pt}}
\put(1456.0,31.0){\rule[-0.200pt]{0.400pt}{4.818pt}}
\put(1456,19){\makebox(0,0){\shortstack{\\ \\ \\ {$5$}}}}
\put(1456.0,873.0){\rule[-0.200pt]{0.400pt}{4.818pt}}
\put(120.0,31.0){\rule[-0.200pt]{321.842pt}{0.400pt}}
\put(1456.0,31.0){\rule[-0.200pt]{0.400pt}{207.656pt}}
\put(120.0,893.0){\rule[-0.200pt]{321.842pt}{0.400pt}}
\put(12,462){\makebox(0,0){{$c_p$}}}
\put(788,-53){\makebox(0,0){{$L.a\surd\sigma$}}}
\put(120.0,31.0){\rule[-0.200pt]{0.400pt}{207.656pt}}
\put(392,272){\makebox(0,0){$+$}}
\put(602,420){\makebox(0,0){$+$}}
\put(812,374){\makebox(0,0){$+$}}
\put(1023,360){\makebox(0,0){$+$}}
\put(1233,377){\makebox(0,0){$+$}}
\put(1443,389){\makebox(0,0){$+$}}
\put(392.0,237.0){\rule[-0.200pt]{0.400pt}{16.863pt}}
\put(382.0,237.0){\rule[-0.200pt]{4.818pt}{0.400pt}}
\put(382.0,307.0){\rule[-0.200pt]{4.818pt}{0.400pt}}
\put(602.0,364.0){\rule[-0.200pt]{0.400pt}{26.981pt}}
\put(592.0,364.0){\rule[-0.200pt]{4.818pt}{0.400pt}}
\put(592.0,476.0){\rule[-0.200pt]{4.818pt}{0.400pt}}
\put(812.0,336.0){\rule[-0.200pt]{0.400pt}{18.549pt}}
\put(802.0,336.0){\rule[-0.200pt]{4.818pt}{0.400pt}}
\put(802.0,413.0){\rule[-0.200pt]{4.818pt}{0.400pt}}
\put(1023.0,342.0){\rule[-0.200pt]{0.400pt}{8.672pt}}
\put(1013.0,342.0){\rule[-0.200pt]{4.818pt}{0.400pt}}
\put(1013.0,378.0){\rule[-0.200pt]{4.818pt}{0.400pt}}
\put(1233.0,358.0){\rule[-0.200pt]{0.400pt}{9.154pt}}
\put(1223.0,358.0){\rule[-0.200pt]{4.818pt}{0.400pt}}
\put(1223.0,396.0){\rule[-0.200pt]{4.818pt}{0.400pt}}
\put(1443.0,379.0){\rule[-0.200pt]{0.400pt}{5.059pt}}
\put(1433.0,379.0){\rule[-0.200pt]{4.818pt}{0.400pt}}
\put(1433.0,400.0){\rule[-0.200pt]{4.818pt}{0.400pt}}
\put(276,429){\makebox(0,0){$\times$}}
\put(428,294){\makebox(0,0){$\times$}}
\put(580,354){\makebox(0,0){$\times$}}
\put(732,403){\makebox(0,0){$\times$}}
\put(884,414){\makebox(0,0){$\times$}}
\put(1036,426){\makebox(0,0){$\times$}}
\put(1188,420){\makebox(0,0){$\times$}}
\put(276.0,212.0){\rule[-0.200pt]{0.400pt}{104.551pt}}
\put(266.0,212.0){\rule[-0.200pt]{4.818pt}{0.400pt}}
\put(266.0,646.0){\rule[-0.200pt]{4.818pt}{0.400pt}}
\put(428.0,230.0){\rule[-0.200pt]{0.400pt}{31.076pt}}
\put(418.0,230.0){\rule[-0.200pt]{4.818pt}{0.400pt}}
\put(418.0,359.0){\rule[-0.200pt]{4.818pt}{0.400pt}}
\put(580.0,314.0){\rule[-0.200pt]{0.400pt}{19.272pt}}
\put(570.0,314.0){\rule[-0.200pt]{4.818pt}{0.400pt}}
\put(570.0,394.0){\rule[-0.200pt]{4.818pt}{0.400pt}}
\put(732.0,369.0){\rule[-0.200pt]{0.400pt}{16.140pt}}
\put(722.0,369.0){\rule[-0.200pt]{4.818pt}{0.400pt}}
\put(722.0,436.0){\rule[-0.200pt]{4.818pt}{0.400pt}}
\put(884.0,398.0){\rule[-0.200pt]{0.400pt}{7.950pt}}
\put(874.0,398.0){\rule[-0.200pt]{4.818pt}{0.400pt}}
\put(874.0,431.0){\rule[-0.200pt]{4.818pt}{0.400pt}}
\put(1036.0,413.0){\rule[-0.200pt]{0.400pt}{6.263pt}}
\put(1026.0,413.0){\rule[-0.200pt]{4.818pt}{0.400pt}}
\put(1026.0,439.0){\rule[-0.200pt]{4.818pt}{0.400pt}}
\put(1188.0,408.0){\rule[-0.200pt]{0.400pt}{6.022pt}}
\put(1178.0,408.0){\rule[-0.200pt]{4.818pt}{0.400pt}}
\put(1178.0,433.0){\rule[-0.200pt]{4.818pt}{0.400pt}}
\put(176,264){\makebox(0,0){$\star$}}
\put(278,634){\makebox(0,0){$\star$}}
\put(380,324){\makebox(0,0){$\star$}}
\put(482,399){\makebox(0,0){$\star$}}
\put(176.0,75.0){\rule[-0.200pt]{0.400pt}{91.060pt}}
\put(166.0,75.0){\rule[-0.200pt]{4.818pt}{0.400pt}}
\put(166.0,453.0){\rule[-0.200pt]{4.818pt}{0.400pt}}
\put(278.0,424.0){\rule[-0.200pt]{0.400pt}{101.178pt}}
\put(268.0,424.0){\rule[-0.200pt]{4.818pt}{0.400pt}}
\put(268.0,844.0){\rule[-0.200pt]{4.818pt}{0.400pt}}
\put(380.0,292.0){\rule[-0.200pt]{0.400pt}{15.418pt}}
\put(370.0,292.0){\rule[-0.200pt]{4.818pt}{0.400pt}}
\put(370.0,356.0){\rule[-0.200pt]{4.818pt}{0.400pt}}
\put(482.0,349.0){\rule[-0.200pt]{0.400pt}{23.849pt}}
\put(472.0,349.0){\rule[-0.200pt]{4.818pt}{0.400pt}}
\put(472.0,448.0){\rule[-0.200pt]{4.818pt}{0.400pt}}
\end{picture}

%% file: plot_break_phys2.tex
\setlength{\unitlength}{0.240900pt}
\ifx\plotpoint\undefined\newsavebox{\plotpoint}\fi
\begin{picture}(1500,900)(0,0)
\font\gnuplot=cmr10 at 12pt
\gnuplot
\sbox{\plotpoint}{\rule[-0.200pt]{0.400pt}{0.400pt}}%
\put(120.0,31.0){\rule[-0.200pt]{4.818pt}{0.400pt}}
\put(108,31){\makebox(0,0)[r]{{$0$}}}
\put(1436.0,31.0){\rule[-0.200pt]{4.818pt}{0.400pt}}
\put(120.0,376.0){\rule[-0.200pt]{4.818pt}{0.400pt}}
\put(108,376){\makebox(0,0)[r]{{$10$}}}
\put(1436.0,376.0){\rule[-0.200pt]{4.818pt}{0.400pt}}
\put(120.0,721.0){\rule[-0.200pt]{4.818pt}{0.400pt}}
\put(108,721){\makebox(0,0)[r]{{$20$}}}
\put(1436.0,721.0){\rule[-0.200pt]{4.818pt}{0.400pt}}
\put(181.0,31.0){\rule[-0.200pt]{0.400pt}{4.818pt}}
\put(181,19){\makebox(0,0){\shortstack{\\ \\ \\ {$1$}}}}
\put(181.0,873.0){\rule[-0.200pt]{0.400pt}{4.818pt}}
\put(484.0,31.0){\rule[-0.200pt]{0.400pt}{4.818pt}}
\put(484,19){\makebox(0,0){\shortstack{\\ \\ \\ {$2$}}}}
\put(484.0,873.0){\rule[-0.200pt]{0.400pt}{4.818pt}}
\put(788.0,31.0){\rule[-0.200pt]{0.400pt}{4.818pt}}
\put(788,19){\makebox(0,0){\shortstack{\\ \\ \\ {$3$}}}}
\put(788.0,873.0){\rule[-0.200pt]{0.400pt}{4.818pt}}
\put(1092.0,31.0){\rule[-0.200pt]{0.400pt}{4.818pt}}
\put(1092,19){\makebox(0,0){\shortstack{\\ \\ \\ {$4$}}}}
\put(1092.0,873.0){\rule[-0.200pt]{0.400pt}{4.818pt}}
\put(1395.0,31.0){\rule[-0.200pt]{0.400pt}{4.818pt}}
\put(1395,19){\makebox(0,0){\shortstack{\\ \\ \\ {$5$}}}}
\put(1395.0,873.0){\rule[-0.200pt]{0.400pt}{4.818pt}}
\put(120.0,31.0){\rule[-0.200pt]{321.842pt}{0.400pt}}
\put(1456.0,31.0){\rule[-0.200pt]{0.400pt}{207.656pt}}
\put(120.0,893.0){\rule[-0.200pt]{321.842pt}{0.400pt}}
\put(12,510){\makebox(0,0){{$l_{\hbox{br}}.a\surd\sigma$}}}
\put(788,-53){\makebox(0,0){$L.a\surd\sigma$}}
\put(120.0,31.0){\rule[-0.200pt]{0.400pt}{207.656pt}}
\put(522,251){\makebox(0,0){$+$}}
\put(737,373){\makebox(0,0){$+$}}
\put(952,495){\makebox(0,0){$+$}}
\put(1167,568){\makebox(0,0){$+$}}
\put(1382,861){\makebox(0,0){$+$}}
\put(522.0,226.0){\rule[-0.200pt]{0.400pt}{11.804pt}}
\put(512.0,226.0){\rule[-0.200pt]{4.818pt}{0.400pt}}
\put(512.0,275.0){\rule[-0.200pt]{4.818pt}{0.400pt}}
\put(737.0,336.0){\rule[-0.200pt]{0.400pt}{17.586pt}}
\put(727.0,336.0){\rule[-0.200pt]{4.818pt}{0.400pt}}
\put(727.0,409.0){\rule[-0.200pt]{4.818pt}{0.400pt}}
\put(952.0,446.0){\rule[-0.200pt]{0.400pt}{23.608pt}}
\put(942.0,446.0){\rule[-0.200pt]{4.818pt}{0.400pt}}
\put(942.0,544.0){\rule[-0.200pt]{4.818pt}{0.400pt}}
\put(1167.0,507.0){\rule[-0.200pt]{0.400pt}{29.390pt}}
\put(1157.0,507.0){\rule[-0.200pt]{4.818pt}{0.400pt}}
\put(1157.0,629.0){\rule[-0.200pt]{4.818pt}{0.400pt}}
\put(1382.0,775.0){\rule[-0.200pt]{0.400pt}{28.426pt}}
\put(1372.0,775.0){\rule[-0.200pt]{4.818pt}{0.400pt}}
\put(1372.0,893.0){\rule[-0.200pt]{4.818pt}{0.400pt}}
\put(343,208){\makebox(0,0){$\times$}}
\put(499,278){\makebox(0,0){$\times$}}
\put(654,331){\makebox(0,0){$\times$}}
\put(810,455){\makebox(0,0){$\times$}}
\put(965,543){\makebox(0,0){$\times$}}
\put(1121,543){\makebox(0,0){$\times$}}
\put(343.0,190.0){\rule[-0.200pt]{0.400pt}{8.431pt}}
\put(333.0,190.0){\rule[-0.200pt]{4.818pt}{0.400pt}}
\put(333.0,225.0){\rule[-0.200pt]{4.818pt}{0.400pt}}
\put(499.0,252.0){\rule[-0.200pt]{0.400pt}{12.768pt}}
\put(489.0,252.0){\rule[-0.200pt]{4.818pt}{0.400pt}}
\put(489.0,305.0){\rule[-0.200pt]{4.818pt}{0.400pt}}
\put(654.0,304.0){\rule[-0.200pt]{0.400pt}{13.009pt}}
\put(644.0,304.0){\rule[-0.200pt]{4.818pt}{0.400pt}}
\put(644.0,358.0){\rule[-0.200pt]{4.818pt}{0.400pt}}
\put(810.0,411.0){\rule[-0.200pt]{0.400pt}{21.199pt}}
\put(800.0,411.0){\rule[-0.200pt]{4.818pt}{0.400pt}}
\put(800.0,499.0){\rule[-0.200pt]{4.818pt}{0.400pt}}
\put(965.0,490.0){\rule[-0.200pt]{0.400pt}{25.535pt}}
\put(955.0,490.0){\rule[-0.200pt]{4.818pt}{0.400pt}}
\put(955.0,596.0){\rule[-0.200pt]{4.818pt}{0.400pt}}
\put(1121.0,490.0){\rule[-0.200pt]{0.400pt}{25.535pt}}
\put(1111.0,490.0){\rule[-0.200pt]{4.818pt}{0.400pt}}
\put(1111.0,596.0){\rule[-0.200pt]{4.818pt}{0.400pt}}
\put(190,138){\makebox(0,0){$\star$}}
\put(295,197){\makebox(0,0){$\star$}}
\put(399,221){\makebox(0,0){$\star$}}
\put(190.0,126.0){\rule[-0.200pt]{0.400pt}{5.782pt}}
\put(180.0,126.0){\rule[-0.200pt]{4.818pt}{0.400pt}}
\put(180.0,150.0){\rule[-0.200pt]{4.818pt}{0.400pt}}
\put(295.0,179.0){\rule[-0.200pt]{0.400pt}{8.672pt}}
\put(285.0,179.0){\rule[-0.200pt]{4.818pt}{0.400pt}}
\put(285.0,215.0){\rule[-0.200pt]{4.818pt}{0.400pt}}
\put(399.0,203.0){\rule[-0.200pt]{0.400pt}{8.672pt}}
\put(389.0,203.0){\rule[-0.200pt]{4.818pt}{0.400pt}}
\put(389.0,239.0){\rule[-0.200pt]{4.818pt}{0.400pt}}
\sbox{\plotpoint}{\rule[-0.500pt]{1.000pt}{1.000pt}}%
\put(120,101){\usebox{\plotpoint}}
\put(120.00,101.00){\usebox{\plotpoint}}
\put(138.92,109.54){\usebox{\plotpoint}}
\put(157.53,118.67){\usebox{\plotpoint}}
\multiput(160,120)(19.077,8.176){0}{\usebox{\plotpoint}}
\put(176.47,127.14){\usebox{\plotpoint}}
\put(195.42,135.61){\usebox{\plotpoint}}
\put(213.93,144.96){\usebox{\plotpoint}}
\multiput(214,145)(19.077,8.176){0}{\usebox{\plotpoint}}
\put(232.94,153.28){\usebox{\plotpoint}}
\put(251.63,162.31){\usebox{\plotpoint}}
\multiput(255,164)(18.845,8.698){0}{\usebox{\plotpoint}}
\put(270.45,171.05){\usebox{\plotpoint}}
\put(289.43,179.43){\usebox{\plotpoint}}
\put(308.08,188.54){\usebox{\plotpoint}}
\multiput(309,189)(18.845,8.698){0}{\usebox{\plotpoint}}
\put(326.97,197.13){\usebox{\plotpoint}}
\put(345.63,206.18){\usebox{\plotpoint}}
\multiput(349,208)(19.077,8.176){0}{\usebox{\plotpoint}}
\put(364.54,214.71){\usebox{\plotpoint}}
\put(383.27,223.64){\usebox{\plotpoint}}
\put(402.02,232.55){\usebox{\plotpoint}}
\multiput(403,233)(19.077,8.176){0}{\usebox{\plotpoint}}
\put(421.03,240.86){\usebox{\plotpoint}}
\put(439.73,249.87){\usebox{\plotpoint}}
\multiput(444,252)(18.845,8.698){0}{\usebox{\plotpoint}}
\put(458.53,258.66){\usebox{\plotpoint}}
\put(477.33,267.41){\usebox{\plotpoint}}
\put(496.11,276.19){\usebox{\plotpoint}}
\multiput(498,277)(18.845,8.698){0}{\usebox{\plotpoint}}
\put(515.03,284.73){\usebox{\plotpoint}}
\put(533.72,293.70){\usebox{\plotpoint}}
\multiput(538,296)(19.077,8.176){0}{\usebox{\plotpoint}}
\put(552.61,302.28){\usebox{\plotpoint}}
\put(571.36,311.18){\usebox{\plotpoint}}
\put(590.08,320.12){\usebox{\plotpoint}}
\multiput(592,321)(19.077,8.176){0}{\usebox{\plotpoint}}
\put(609.01,328.62){\usebox{\plotpoint}}
\put(627.64,337.70){\usebox{\plotpoint}}
\multiput(633,340)(18.845,8.698){0}{\usebox{\plotpoint}}
\put(646.56,346.24){\usebox{\plotpoint}}
\put(665.40,354.91){\usebox{\plotpoint}}
\put(684.15,363.78){\usebox{\plotpoint}}
\multiput(687,365)(18.845,8.698){0}{\usebox{\plotpoint}}
\put(702.98,372.49){\usebox{\plotpoint}}
\put(721.66,381.53){\usebox{\plotpoint}}
\put(740.67,389.86){\usebox{\plotpoint}}
\multiput(741,390)(18.275,9.840){0}{\usebox{\plotpoint}}
\put(759.18,399.22){\usebox{\plotpoint}}
\put(778.13,407.68){\usebox{\plotpoint}}
\multiput(781,409)(19.077,8.176){0}{\usebox{\plotpoint}}
\put(797.08,416.12){\usebox{\plotpoint}}
\put(815.68,425.29){\usebox{\plotpoint}}
\put(834.60,433.82){\usebox{\plotpoint}}
\multiput(835,434)(18.564,9.282){0}{\usebox{\plotpoint}}
\put(853.23,442.95){\usebox{\plotpoint}}
\put(872.20,451.37){\usebox{\plotpoint}}
\multiput(876,453)(18.845,8.698){0}{\usebox{\plotpoint}}
\put(891.06,460.03){\usebox{\plotpoint}}
\put(909.73,469.10){\usebox{\plotpoint}}
\put(928.73,477.45){\usebox{\plotpoint}}
\multiput(930,478)(18.275,9.840){0}{\usebox{\plotpoint}}
\put(947.23,486.81){\usebox{\plotpoint}}
\put(966.20,495.24){\usebox{\plotpoint}}
\multiput(970,497)(18.564,9.282){0}{\usebox{\plotpoint}}
\put(984.83,504.38){\usebox{\plotpoint}}
\put(1003.76,512.90){\usebox{\plotpoint}}
\put(1022.69,521.40){\usebox{\plotpoint}}
\multiput(1024,522)(18.564,9.282){0}{\usebox{\plotpoint}}
\put(1041.32,530.53){\usebox{\plotpoint}}
\put(1060.28,538.98){\usebox{\plotpoint}}
\multiput(1065,541)(18.275,9.840){0}{\usebox{\plotpoint}}
\put(1078.79,548.34){\usebox{\plotpoint}}
\put(1097.80,556.67){\usebox{\plotpoint}}
\put(1116.78,565.05){\usebox{\plotpoint}}
\multiput(1119,566)(18.275,9.840){0}{\usebox{\plotpoint}}
\put(1135.29,574.41){\usebox{\plotpoint}}
\put(1154.27,582.81){\usebox{\plotpoint}}
\put(1172.90,591.95){\usebox{\plotpoint}}
\multiput(1173,592)(18.845,8.698){0}{\usebox{\plotpoint}}
\put(1191.81,600.49){\usebox{\plotpoint}}
\put(1210.43,609.62){\usebox{\plotpoint}}
\multiput(1213,611)(19.077,8.176){0}{\usebox{\plotpoint}}
\put(1229.37,618.09){\usebox{\plotpoint}}
\put(1248.32,626.56){\usebox{\plotpoint}}
\put(1266.83,635.91){\usebox{\plotpoint}}
\multiput(1267,636)(19.077,8.176){0}{\usebox{\plotpoint}}
\put(1285.84,644.23){\usebox{\plotpoint}}
\put(1304.53,653.26){\usebox{\plotpoint}}
\multiput(1308,655)(18.845,8.698){0}{\usebox{\plotpoint}}
\put(1323.35,662.01){\usebox{\plotpoint}}
\put(1342.33,670.38){\usebox{\plotpoint}}
\put(1360.98,679.49){\usebox{\plotpoint}}
\multiput(1362,680)(18.845,8.698){0}{\usebox{\plotpoint}}
\put(1379.87,688.09){\usebox{\plotpoint}}
\put(1398.53,697.13){\usebox{\plotpoint}}
\multiput(1402,699)(19.077,8.176){0}{\usebox{\plotpoint}}
\put(1417.44,705.66){\usebox{\plotpoint}}
\put(1436.17,714.59){\usebox{\plotpoint}}
\put(1454.92,723.50){\usebox{\plotpoint}}
\put(1456,724){\usebox{\plotpoint}}
\end{picture}